\documentclass[11pt]{amsart}
\usepackage[T1]{fontenc}
\usepackage{a4wide}
\usepackage{mathpazo}
\usepackage{mathrsfs}
\usepackage{amssymb, amsmath}
\usepackage{enumitem}
\usepackage{booktabs}
\usepackage{array}
\usepackage{xcolor}
\usepackage{cite}
\usepackage{hyperref}
\hypersetup{colorlinks=true, linkcolor=blue!50!black, citecolor=green!35!black, urlcolor=blue!50!black}

\newcommand{\acc}{\mathit{acc}}
\newcommand{\SchedCost}{C}
\newcommand{\thr}{\mathrm{thr}}
\newcommand{\SchedCostP}{C'}
\newcommand{\N}{\mathbb{N}}

\theoremstyle{plain}
\newtheorem{thm}{Theorem}
\newtheorem{lem}[thm]{Lemma}
\newtheorem{prop}[thm]{Proposition}

\theoremstyle{definition}
\newtheorem{defn}[thm]{Definition}
\theoremstyle{remark}
\newtheorem{rem}[thm]{Remark}

\begin{document}

\title[A sharp additive bound for the threshold rule]{A Sharp Additive Bound for the Threshold Rule\\
in Repeated Rent-or-Buy}

\begin{abstract}
Motivated by context management in LLM agents, we study a repeated rent-or-buy model with schedule-independent
history growth and a fixed reset price $K$, a discrete form of dynamic TCP acknowledgment:
every call pays as rent the history accumulated since the last reset, and a reset clears it.
The classical threshold rule is $2$-competitive, and no deterministic rule guarantees a smaller
ratio. We show that, measured additively, its loss is governed by the changes in the growth: if
the growth is constant on $P$ runs, the threshold rule, knowing neither the runs nor the
horizon, pays at most $(2P-1)K$ more than the optimal schedule chosen with hindsight, and the
coefficient $2P-1$ is sharp. If the growth is only within a factor $1+\varepsilon$ of $P$
runs, the bound holds up to a factor $1+\varepsilon$ and an extra $\varepsilon K$. The proof is a potential argument built from the relative
value functions of the single-rate problems. On two runs no deterministic rule guarantees a
loss below $2K$. A rule that also resets when the growth decreases guarantees $2K$, while the
threshold rule can lose arbitrarily close to $3K$.
\end{abstract}

\author{Vasco Schiavo}
\email{vasco.schiavo@protonmail.com}

\keywords{Online algorithms, rent-or-buy, dynamic TCP acknowledgment, additive guarantees,
competitive analysis, context management, LLM agents}

\maketitle

\section{Introduction}\label{sec:intro}

A session makes calls $t=1,\dots,T$. At call $t$, a schedule observes the history $h_t$
and decides whether to \emph{reset}. The call pays $h_t$ as \emph{rent} in either case.
A reset also costs $K>0$ and clears that history. Each call $t<T$ then appends an increment
$a_t\ge0$ for the next call. Neither the increments nor the horizon $T$ are known in advance.
This is a repeated rent-or-buy problem, a finite-horizon discrete variant of dynamic TCP acknowledgment
\cite{dooly2001tcp, karlin2003tcp}. It models an LLM agent, which re-sends its whole history
on every call and is billed for it \cite{weinberger2026token, spend2026, cache2026}, and can
summarize the history at a fixed price: when should it summarize? The model assumes that the
increments are the same under every schedule, that $K$ is fixed, and that a reset
clears the modeled history and changes nothing the agent does afterwards. A part re-sent by
every call, such as the system prompt, is common to all schedules and left out. The model
leaves out the effect of a summary on the information the agent keeps and on a prompt cache,
and summaries whose size depends on what they summarize. So the theorems bound the modeled
cost only.

\subsection*{What is known.}
The classical \emph{threshold rule} accumulates the rent paid on non-reset calls since its
last reset. It resets when adding the current rent would bring the accumulator to at least
$K$. It pays at most twice the cost of the best schedule chosen with hindsight, and no
deterministic rule guarantees a smaller ratio (Section~\ref{sec:model}), as in the classical
TCP problem \cite[Theorems~10 and~12]{dooly1998tr}. For standard dynamic TCP acknowledgment,
randomized algorithms achieve $\frac{e}{e-1}$ \cite[Theorem~1]{karlin2003tcp}. Our model differs
in that terminal history need not be cleared. Under constant growth the rule resets with the
period that minimizes the cost per call, a discrete form of Harris's economic order quantity
\cite{harris1913eoq}. The factor $2$ is a worst case over all sessions, attained by a single
burst of growth followed by nothing, with the session ending right after the rule resets. It
does not say how the loss depends on the session.

\subsection*{Our result.}
A \emph{run} is a maximal block of equal consecutive increments, and $P$ is the number of
runs. We show that the loss of the threshold rule, which does not know the runs, their number
or the horizon, is governed by $P$ alone, whatever the length of the session or the size of
the increments.

\begin{thm}[Main Theorem]\label{thm:phase}
Let $K>0$ and $P\in\N_{\ge1}$.
\begin{enumerate}[nosep]
\item For every session $S$ with $P$ runs, of arbitrary nonnegative
      rates, the threshold rule satisfies
      \[
       \SchedCost_{r^{\thr}}(S)\;\le\;\min\bigl\{\,2\,\SchedCost_{r^\star}(S),\;\SchedCost_{r^\star}(S)+(2P-1)K\,\bigr\}.
      \]
\item For every $\eta>0$ there is a session $S$ with exactly $P$ runs such that
    \begin{align*}
        \SchedCost_{r^{\thr}}(S)>\SchedCost_{r^\star}(S)+(2P-1-\eta)K.
    \end{align*}
\end{enumerate}
\end{thm}

In words, the threshold rule pays one reset price for not knowing the session and two for
each change of growth. The first cannot be improved by any rule (Remark~\ref{rem:base}), and
part~(2) shows that the second is sharp. When ${(2P-1)K}<\SchedCost_{r^\star}(S)$, the bound
also improves the ratio $2$ to $1+\frac{(2P-1)K}{\SchedCost_{r^\star}(S)}$.

\medskip
The rule is not new: the commit rule of CADOC, which times context replacement, is of this
kind \cite{cadoc2026}. What is new is the guarantee: a sharp additive bound against the
unrestricted optimum, with no predictions or stochastic assumptions, and the best
deterministic loss on two runs.

\medskip
The parameter $P$ counts exact changes. If every
increment is within a factor $1+\varepsilon$ of a session with $P$ runs, the rule pays at most
\begin{align*}
    (1+\varepsilon)\bigl(\SchedCost_{r^\star}(S)+(2P-1)K\bigr)+\varepsilon K
\end{align*}
by Remark~\ref{rem:approx}.

\subsection*{The proof.}
The runs cannot be analyzed one at a time: neither the rule nor the optimal schedule restarts
at a boundary, and both carry rent across it. We compare the schedules using potentials
built from the single-rate problems. Within each run, these potentials bound the excess
cost by a telescoping sum. At a change of rate, the old and new potential differences lie
in the same interval of width $2K$, so the change contributes at most $2K$. The terminal
term contributes at most $K$, giving $K+2(P-1)K$.

\subsection*{The rule or the problem?}
On sessions with at most two runs we can say which part of the loss is the rule's fault. No
deterministic rule guarantees a loss below $2K$ (Theorem~\ref{thm:lower2}), since none can tell
slow growth that goes on from slow growth about to be followed by a burst. The threshold rule
loses at most $2K$ when the change is an increase, but arbitrarily close to $3K$ at a decrease,
where it keeps paying the stale rent of the fast run. All the sharp sessions of
Section~\ref{sec:sharp} change by decreases. A rule that also resets at every decrease loses
at most $K$ per decrease (Theorem~\ref{thm:drop}) and $2K$ on two runs, the best possible
(Theorem~\ref{thm:rise}), though it gives up the factor $2$: it is not $c$-competitive for any
constant $c$. Table~\ref{tab:price} records these results as prices: a rule $\pi$ has prices
$(b,d,u)\in[0,\infty)^3$ if every session $S$ satisfies
$\SchedCost_{r^\pi}(S)\le\SchedCost_{r^\star}(S)+(b+dP_-+uP_+)K$, where $P_-$ and $P_+$ count
the decreases and the increases between consecutive runs. Prices are coefficients of a
guarantee for all sessions, not a split of the loss on one. Whether some rule pays $K$ for
every change, on any number of runs, is open.

\begin{table}[h]
\centering\small
\caption{Prices $(b,d,u)$ in units of $K$. Larger triples are prices too. The first two
rows fix the least base, $b=1$, and give the $d$ and $u$ we prove with it and the known lower
bounds. The last row gives limits for every rule. On sessions with two runs, $u=1$ suffices
for both rules, and no rule with $b=1$ has a smaller $u$ (Theorems~\ref{thm:lower2}
and~\ref{thm:rise}).}\label{tab:price}
\begin{tabular}{@{}llll@{}}
\toprule
 & base $b$ & decrease $d$ & increase $u$\\
\midrule
threshold rule  & $1$ & $2$, sharp (Sec.~\ref{sec:sharp}) & between $1$ and $2$\\
drop-aware rule & $1$ & at most $1$ (Thm.~\ref{thm:drop}) & between $1$ and $2$\\
every rule      & at least $1$ (Rem.~\ref{rem:base}) & open & $b+u\ge2$ (Thm.~\ref{thm:lower2})\\
\bottomrule
\end{tabular}
\end{table}

\subsection*{Related work.}
Beyond the worst case, guarantees exist for predicted resets \cite{purohit2018ml,
bamas2020primaldual, im2023ack, angelopoulos2019advice}, for general delay costs
\cite{bhore2026delay}, for stationary stochastic arrivals up to constant factors
\cite[Section~3]{mari2024aggregation}, and against the best of $k$ online experts: Harada,
Kawase and Sumita \cite{harada2026tcp} give a $(2+\varepsilon)$-competitive randomized rule with
expected regret $O(\sqrt{\mathrm{OPT}\log(k+1)}+\log(\frac{1}{\varepsilon}))$ against the best of them,
in the tension between ratio and regret studied by Andrew et al.\ \cite{andrew2013tale}.
Closest in spirit is tracking regret, which grows with the number of switches of the
comparator \cite{herbster1998tracking}. Here, the changes are in the input instead, and the comparator
is the unrestricted optimum. Worst-case analyses of lot-sizing heuristics bound ratios to the
optimum \cite{axsater1982worst, vachani1992heuristics}, which no heuristic that ignores future
demand brings below $2$ \cite{axsater1985bounds, heuvel2010online}. Vachani refines them by the
horizon and for constant or bounded demand. These are ratios, and none of them depends on the
number of changes in the demand. For agents, CADOC batches context replacements with
a stationary-optimal batch size and a commit rule that is $2$-competitive for its
waiting-plus-setup cost when the setup cost is a positive constant
\cite[Propositions~1 and~2]{cadoc2026}. Other compaction timings are tuned or learned
\cite{cliff2026, autocompact2026, selfgc2026}. Section~\ref{sec:model} sets up the model,
Sections~\ref{sec:phase} and~\ref{sec:sharp} prove Theorem~\ref{thm:phase}, and
Section~\ref{sec:every} compares rules.

\medskip
\subsection*{Disclosure}
A large language model was used to help prepare this paper. The author has read and checked
every result and takes full responsibility for the paper, including any errors.

\section{Model and classical facts}\label{sec:model}

We write $\N=\{0,1,2,\dots\}$ and $\N_{\ge1}=\N\setminus\{0\}$.

\begin{defn}[Session]\label{def:session}
Let $T\in\N_{\ge1}$. A \textbf{session} $S$ of \textbf{horizon} $T$ is a sequence
$(a_t)_{t=1}^{T-1}\in[0,\infty)^{T-1}$: it has $T$ calls, $a_t$ is the \textbf{increment}
that call $t$ appends to the history, and the sequence of increments is the \textbf{growth} of
$S$. A
\textbf{run} of $S$ is a maximal set of consecutive indices $I\subseteq\{1,\dots,T-1\}$ such that $a_i = a_j$ for every $i, j \in I$, and this constant is the \textbf{rate} of the run.
\end{defn}

The number of runs $P$ of a session $S$ of horizon $T$ is
\begin{align*}
    P = \left\{
            \begin{array}{ll}
                0 & \quad \text{if } T = 1 \\
                1+\bigl|\{t\in\{2,\dots,T-1\}:a_t\ne a_{t-1}\}\bigr|  &\quad \text{otherwise.}
            \end{array}
        \right.
\end{align*}

From now on, we suppose that $T\ge2$, so $P\ge1$.

\begin{defn}[Schedule and cost]\label{def:cost}
Let $S$ be a session of horizon $T$, and let $K>0$ be the \textbf{price} of a reset, fixed
throughout. A \textbf{schedule} for $S$ is an element $r\in\{0,1\}^T$, where
$r_t=1$ means reset before call $t$. Its \textbf{history}, the \textbf{cost of call $t$} and
the \textbf{total cost} are
\[
 h_1=0,\quad h_{t+1}=(1-r_t)\,h_t+a_t\ \ (1\le t<T),\qquad
 c_t(r)=r_tK+h_t,\qquad \SchedCost_r(S)=\sum_{t=1}^{T}c_t(r).
\]
We suppress the dependence of $h_t$ on the schedule $r$.
\end{defn}

Since $\{0,1\}^T$ is finite, there is a schedule $r^\star$ for $S$, possibly not
unique, with $\SchedCost_{r^\star}(S)\le\SchedCost_r(S)$ for every schedule $r$ for $S$. We call
such a schedule optimal with hindsight.

\medskip
At call $t$, the schedule observes $h_t$ and chooses $r_t$. It pays $h_t+r_tK$, clears
the history if $r_t=1$, and then appends $a_t$ if $t<T$. The rent includes $h_t$ even on
a reset call because the reset reads the history it clears. There is no required reset
after the final call and no cost beyond $T$. For example, with increments $a_1=x$ and
$a_2=y$ the three calls cost $0$, $x$ and $x+y$ without a reset, and $0$, $x+K$ and $y$ with a
reset before call~$2$. An increment appended at call $t$ and cleared by a reset before call $t'$ thus
pays $(t'-t)\,a_t$, and an increment never cleared pays $(T-t)\,a_t$.

\medskip
In the language of dynamic TCP acknowledgment with the objective $f_{\mathrm{sum}}$ of
\cite[Definition~2]{dooly1998tr}, under $\SchedCost$ every increment waits at least one step:
increments arrive just after the integer times and acknowledgments happen at the integers.
The other discretization, the cost $\SchedCostP$ of Remark~\ref{rem:other}, is the
convention in which an acknowledgment also covers the arrivals of its own step, so a reset
call pays the price only. The additive bound of Theorem~\ref{thm:phase} holds and is sharp
under both (Remark~\ref{rem:other}).

\begin{defn}[Rule]\label{def:rule}
A \textbf{rule} is a map $\pi\colon\bigcup_{n\ge0}[0,\infty)^n\to\{0,1\}$, where
$[0,\infty)^0$ contains only the empty sequence. On a session $S$ of horizon $T$ it
produces the schedule $r^\pi$ given by $r^\pi_t=\pi(a_1,\dots,a_{t-1})$ for $t=1,\dots,T$.
\end{defn}

Rules are thus deterministic and do not know the horizon: on a session that extends $S$ by
further calls, a rule produces a schedule whose first $T$ entries are the ones it produces
on $S$.

\medskip
A rule $\pi$ \textbf{loses} $\SchedCost_{r^\pi}(S)-\SchedCost_r(S)$ against a schedule $r$ for $S$, and its
\textbf{loss} on $S$ is $\SchedCost_{r^\pi}(S)-\SchedCost_{r^\star}(S)$, the most it loses
against any schedule. We call $\pi$ \textbf{$c$-competitive} if
$\SchedCost_{r^\pi}(S)\le c\,\SchedCost_{r^\star}(S)$ for every session $S$, with no additive allowance.

\begin{defn}[Threshold rule]\label{def:thr}
On a session $S$ of horizon $T$, let $h_t$ be the history of the schedule being built and
$\acc_t$ the sum of rents paid on non-reset calls since its last reset.
Starting from $\acc_1=0$, for $t=1,\dots,T$
\[
(r_t,\ \acc_{t+1})=
\begin{cases}
(1,\ 0) & \text{if } \acc_t+h_t\ge K,\\
(0,\ \acc_t+h_t) & \text{otherwise.}
\end{cases}
\]
We call $\acc_t+h_t$ the \textbf{test value} of call $t$. Since $h_t$ and $\acc_t$ depend
only on $a_1,\dots,a_{t-1}$, this is a rule, the \textbf{threshold rule}, and its schedule on $S$
is the \textbf{threshold schedule} $r^{\thr}$.
\end{defn}

For a rate $\alpha>0$ we define the two functions
\begin{alignat*}{2}
 g_\alpha&\colon\N\longrightarrow[0,\infty), &\qquad
 n&\longmapsto g_\alpha(n)=\frac{\alpha\,n(n+1)}{2},\\
 f_\alpha&\colon\N_{\ge1}\longrightarrow(0,\infty), &\qquad
 n&\longmapsto f_\alpha(n)=\frac{K+g_\alpha(n)}{n}.
\end{alignat*}
Under constant growth $a_t=\alpha$, the function $g_\alpha(n)$ gives the rent of the first
$n$ calls after a reset, and $f_\alpha(n)$ gives the cost per call of resetting every $n$ calls.
Define
\[
 L_\alpha=\min\{n\in\N_{\ge1} : g_\alpha(n)\ge K\}
 \quad\text{and}\quad f^\ast_\alpha=f_\alpha(L_\alpha).
\]
The minimum exists because $g_\alpha$ is unbounded. Since
$g_\alpha(L_\alpha-1)<K$, a direct computation gives $\alpha L_\alpha\le f^\ast_\alpha$.

\begin{prop}\label{prop:classical}
The following hold:
\begin{enumerate}[nosep]
\item $\SchedCost_{r^{\thr}}(S)\le2\,\SchedCost_r(S)$ for every session $S$ and every
      schedule $r$ for $S$.
\item For every $\alpha>0$, $f^\ast_\alpha=\min_{n\in\N_{\ge1}}f_\alpha(n)$.
\end{enumerate}
\end{prop}

\begin{proof}
Part (1) is the discrete analogue of the factor two of the rule that Dooly, Goldman and
Scott call greedy$_{\mathrm{new}}$ \cite[Theorem~10]{dooly1998tr}. We give the proof in this
model. 

Split the calls into the cycles of the rule: a cycle ends with a call at which the
rule resets, or with call $T$. Both costs are sums over calls, so it suffices to show that
on each cycle the rule pays at most twice what a schedule $r$ pays. Every schedule pays each
increment $a_u$ at call $u+1$, so $r$ pays at least the total $H$ of the increments first
billed in the cycle. The rule's history was empty before the first of them was appended,
so its history at each call of the cycle is at most that of $r$ if $r$ does not reset in the
cycle, and $H$ is the rule's rent at the last call of the cycle.

If the cycle ends with a reset, let $\acc$ be the rule's accumulator at that call. Then
$\acc<K\le\acc+H$, and the rule pays $K+\acc+H$ on the cycle. If $r$ resets in the cycle,
it pays at least $K+H$, and $K+\acc+H<2K+H\le2(K+H)$. If it does not, it pays at least the
rule's rents, $\acc+H$, and $K+\acc+H\le2(\acc+H)$. If the cycle ends with call $T$ and no
reset, the rule pays only its rents, which total less than $K$. Then $r$ pays at least $K$
if it resets in the cycle and at least the rule's rents otherwise.

Part (2) is a discrete counterpart of Harris's economic order quantity
\cite{harris1913eoq}. Directly,
\begin{align*}
    f_\alpha(n+1)-f_\alpha(n)=\frac{g_\alpha(n)-K}{n(n+1)}.
\end{align*}
Since $g_\alpha$ is strictly increasing and $L_\alpha$ is the first $n$ with
$g_\alpha(n)\ge K$, this difference is negative for $n<L_\alpha$, nonnegative for
$n=L_\alpha$ and positive for $n>L_\alpha$.
\end{proof}

The factor $2$ in part~(1) is best possible for deterministic rules, by the adversary of ski
rental \cite{karlin1988snoopy}, used for acknowledgment in \cite[Theorem~12]{dooly1998tr}.
Append a single increment $\delta>0$, followed by zeros, and end the session right after
the rule's first reset, or sufficiently late if it never resets. If it resets before call~$1$,
end after call~$2$. The rule then pays at least $2-\frac{\delta}{K+\delta}$ times the optimum.
Applied to the threshold rule itself, this adversary stops at its first reset: the rule
then loses exactly $K$, against an optimum below $K+\delta$. The ratio $2$ thus comes
from sessions whose optimum is about $K$. On longer sessions it allows a loss as large as the
optimum, while Theorem~\ref{thm:phase} bounds the loss by $(2P-1)K$.

\section{The upper bound}\label{sec:phase}

In this section we prove point (1) of Theorem~\ref{thm:phase}.

\medskip
Without loss of generality, we may assume that every increment is
positive. Indeed, replace each zero increment of $S$ by some $\delta>0$ smaller than every
positive increment of $S$: the runs and the direction of each change stay the same, and the
cost of every schedule is affine in $\delta$. Along the decisions that the threshold rule makes
on $S$, its test values are affine and nondecreasing in $\delta$, so for $\delta$ small enough a
test that holds still holds and one that fails still fails: the rule makes the same decisions.
Letting $\delta\to0$ carries the bound back to $S$. The same argument applies to the other
upper bounds below and to the drop-aware rule of Section~\ref{sec:drop}, whose comparisons
between increments do not change.

\medskip
For $2\le t\le T+1$, the \textbf{carry} of a schedule $r$ before call $t$ is the history left
after the preceding call's reset decision:
\[
 o_t=(1-r_{t-1})\,h_{t-1}\ge0.
\]
For $t\le T$, the new increment
$a_{t-1}$ is added to this carry, so $h_t=o_t+a_{t-1}$. At $t=T+1$, the carry is what the
schedule leaves after its last call. We write $e_t$ for the threshold rule's carry, and use
$e$ and $o$ for the two carries at a fixed call.

\medskip
The \textbf{state} of the threshold rule before call $t$ is $(e_t,\acc_t)$. The carry records
retained history. The accumulator records rent already paid on non-reset calls, so the same
increment can contribute to it repeatedly. At a call with rate $\alpha=a_{t-1}$, the rent is
$e+\alpha$. If $\acc+e+\alpha\ge K$, the rule also pays $K$ and moves to $(0,0)$. Otherwise,
it moves to $(e+\alpha,\acc+e+\alpha)$. By induction, $0\le e\le\acc<K$ at every state it visits.

\medskip
Fix a rate $\alpha>0$. To compare the schedules, we consider the cost of their next reset
if the rate stays $\alpha$, measured relative to the optimal cycle average $f^\ast_\alpha$
per call. For a carry $e\ge0$ and $j\in\N$, define
\[
 R^\alpha_j(e)=g_\alpha(j)+je,\qquad V^\alpha_j(e)=K+R^\alpha_j(e)-jf^\ast_\alpha,\qquad
 W_\alpha(e)=\min_{j\ge0}V^\alpha_j(e).
\]
For $j\ge1$, $R^\alpha_j(e)$ is the rent up to a reset at the $j$-th call: those calls pay
$e+\alpha,\dots,e+j\alpha$. Adding the reset price and subtracting $jf^\ast_\alpha$ gives
$V^\alpha_j(e)$. Minimizing over $j$ gives the benchmark $W_\alpha(e)$. We also include the
artificial candidate $V^\alpha_0(e)=K$ to cap $W_\alpha$ at $K$. This is not a reset action in
the model. The minimum exists because $V^\alpha_j(e)\ge g_\alpha(j)-jf^\ast_\alpha\to\infty$.

\medskip
For the threshold rule, the state $(e,\acc)$ fixes the next reset under constant rate
$\alpha$. Its \textbf{crossing index} is the first $k\ge1$ with
$\acc+R^\alpha_k(e)\ge K$, and its relative cost is $D_\alpha(e,\acc)=V^\alpha_k(e)$.
The proof compares this cost with the benchmark at the comparator's carry through
$Z=D_\alpha(e,\acc)-W_\alpha(o)$. The comparator need not choose a minimizing reset:
$W_\alpha$ gives an inequality for either of its actions (Lemma~\ref{lem:step}). Summing
these one-call inequalities within each run cancels the intermediate potentials, leaving
only the rate changes and the terminal state to be charged.

\begin{lem}\label{lem:rate}
Let $\alpha>0$ and $e\ge0$.
\begin{enumerate}[nosep]
\item For $j\ge1$, $V^\alpha_j(e)=j\bigl(f_\alpha(j)-f^\ast_\alpha\bigr)+je$. In particular
      $V^\alpha_j(e)\ge je$ for all $j$, $D_\alpha(e,\acc)\ge e$ for every $\acc\ge0$, and
      $D_\alpha(0,0)=V^\alpha_{L_\alpha}(0)=0$.
\item $W_\alpha(e)\in[0,K]$, and $W_\alpha(0)=0$.
\item The equality $V^\alpha_{j+1}(e)= V^\alpha_j(e+\alpha) + e+\alpha-f^\ast_\alpha$ holds. Consequently,
    \begin{align*}
        W_\alpha(e)\le W_\alpha(e+\alpha) + e+\alpha-f^\ast_\alpha
        \quad\text{ and }\quad 
        W_\alpha(e)\le K+e+\alpha-f^\ast_\alpha.
    \end{align*}
\item If $e\le\acc<K$ then $0\le D_\alpha(e,\acc)-W_\alpha(e)\le K$.
\item The map $e\mapsto W_\alpha(e)$ is nondecreasing on $[0,\infty)$.
\end{enumerate}
\end{lem}

\begin{proof}
(1) For $j\ge1$,
    \begin{align*}
        V_j^{\alpha}(e) & = K + R_j^\alpha (e) - j f_\alpha^*
            = K + g_\alpha(j) +je - j f_\alpha^* = j\bigl(f_\alpha(j)-f^\ast_\alpha\bigr)+je,
    \end{align*}
where we used in the third equality the fact that $g_\alpha(j) = jf_\alpha(j) - K$.
By Proposition~\ref{prop:classical} point (2), $f_\alpha(j)\ge f^\ast_\alpha$, so
$V^\alpha_j(e)\ge je$, also for $j=0$, where $V^\alpha_0(e)=K$. In addition, $V^\alpha_{L_\alpha}(0)=0$. Hence
\begin{align*}
    D_\alpha(e,\acc)=V^\alpha_k(e)\ge ke\ge e,
\end{align*}
as $k\ge1$, and the crossing index of $(0,0)$ is $L_\alpha$, since $g_\alpha(j)\ge K$ first holds
at $j=L_\alpha$.

(2) Every term is nonnegative by~(1) and the term $j=0$ equals $K$, and
$W_\alpha(0)=V^\alpha_{L_\alpha}(0)=0$.

(3) Expanding the quadratic rents gives
    \begin{align*}
        V_{j + 1}^\alpha(e) & = K + R_{j+1}^\alpha(e) - (j + 1)f_\alpha^* \\
             & = K + g_\alpha(j + 1) + (j + 1) e - (j + 1)f_\alpha^* \\
             & = K + g_\alpha(j) + (j + 1) \alpha + (j+ 1) e - (j + 1)f_\alpha^* \\
             & = K + g_\alpha(j) + j(\alpha + e) - jf_\alpha^* + e + \alpha - f_\alpha^* \\
             & = K + R_j^\alpha(e + \alpha) - jf_\alpha^* + e + \alpha - f_\alpha^* \\
             & = V_j^\alpha(e + \alpha) + e + \alpha - f_\alpha^*, 
    \end{align*}
where the third equality uses $g_\alpha(j+1)-g_\alpha(j)=(j+1)\alpha$. Taking the minimum
over $j\ge0$ on the right gives the first inequality, and $j=0$ gives the second since
$V^\alpha_0(e)=K$.

(4) The lower bound follows from the definition of $W_\alpha$.
For the upper bound we show $V^\alpha_k(e)-V^\alpha_j(e)\le K$ for every $j\ge0$.
Since $\acc+R^\alpha_{L_\alpha}(e)\ge g_\alpha(L_\alpha)\ge K$, the crossing index satisfies
$k\le L_\alpha$. Its minimality also gives $\acc+R^\alpha_{k-1}(e)<K$.

First let $j<k$. Since $R^\alpha_k(e)=R^\alpha_{k-1}(e)+e+k\alpha$,
\begin{align*}
 V^\alpha_k(e)-V^\alpha_j(e)& = R^\alpha_{k-1}(e)-R^\alpha_j(e)+(e+k\alpha-f^\ast_\alpha)-(k-j-1)f^\ast_\alpha \\
 & \le R^\alpha_{k-1}(e)+e \\
 & <K-\acc+e\le K,
\end{align*}
using $R^\alpha_j(e)\ge0$, $k\alpha\le \alpha L_\alpha\le f^\ast_\alpha$ and $e\le\acc$.

Now let $j\ge k$. If $V^\alpha_k(e)\le K$, use $V^\alpha_j(e)\ge0$.
Otherwise, $V^\alpha_k(e)>K$ means $e+\frac{\alpha(k+1)}{2}>f^\ast_\alpha$, and then
\begin{align*}
    V^\alpha_j(e)-V^\alpha_k(e)=(j-k)\bigl(e-f^\ast_\alpha+\frac{\alpha(j+k+1)}{2}\bigr)\ge0.
\end{align*}

(5) For every $j\ge0$ the map $e\mapsto V^\alpha_j(e)=K+g_\alpha(j)+je-jf^\ast_\alpha$ is
nondecreasing, and so is their minimum $W_\alpha$.
\end{proof}

\begin{lem}\label{lem:step}
Let $\alpha>0$, and let $2\le t\le T$ be a call with $a_{t-1}=\alpha$.
\begin{enumerate}[nosep]
\item If $(e,\acc)$ and $(e',\acc')$ are the states of the threshold rule before calls $t$
      and $t+1$, then
      \[c_t(r^{\thr})-f^\ast_\alpha=D_\alpha(e,\acc)-D_\alpha(e',\acc').\]
\item If $o$ and $o'$ are the carries of a schedule $r$ before calls $t$ and $t+1$, then
      \[c_t(r)-f^\ast_\alpha\ge W_\alpha(o)-W_\alpha(o').\]
\end{enumerate}
\end{lem}

\begin{proof}
(1) If the rule does not reset, $k\ge2$ and the crossing index of the next state is
$k-1$, since, for every $n\ge1$,
\begin{align*}
    \acc'+R^\alpha_{n-1}(e')=\acc+e+\alpha+g_\alpha(n-1)+(n-1)(e+\alpha)=\acc+R^\alpha_n(e).
\end{align*}
Hence $D_\alpha(e,\acc)-D_\alpha(e',\acc')=V^\alpha_k(e)-V^\alpha_{k-1}(e+\alpha)=e+\alpha-f^\ast_\alpha$ by
Lemma~\ref{lem:rate} point~(3), which is $c_t(r^{\thr})-f^\ast_\alpha$. If it resets, $k=1$,
$D_\alpha(e,\acc)=V^\alpha_1(e)=K+e+\alpha-f^\ast_\alpha$ equals $c_t(r^{\thr})-f^\ast_\alpha$, and the next potential
is $D_\alpha(0,0)=0$ by Lemma~\ref{lem:rate} point~(1).

(2) If $r_t=0$, then $c_t(r)=o+\alpha$ and $o'=o+\alpha$. If $r_t=1$, then
$c_t(r)=K+o+\alpha$ and $o'=0$. The two inequalities of Lemma~\ref{lem:rate} point~(3), with $W_\alpha(0)=0$ by
Lemma~\ref{lem:rate} point~(2), are exactly
the two cases.
\end{proof}

\begin{proof}[Proof of Theorem~\ref{thm:phase} point~(1)]
Fix a schedule $r$ for $S$. For $2\le t\le T+1$ let $(e_t,\acc_t)$ be the state of the
threshold rule and $o_t$ the carry of $r$ before call $t$, and for a rate $\alpha>0$ put
\[
 Z^\alpha_t=D_\alpha(e_t,\acc_t)-W_\alpha(o_t).
\]
For $1\le t\le T$ let $\Delta_t=c_t(r^{\thr})-c_t(r)$ be the excess cost of call $t$, and
for $2\le t\le T$ write $\rho_t=a_{t-1}$ for the rate of call $t$.

We first bound the excess cost of one call. For $2\le t\le T$, Lemma~\ref{lem:step} points~(1)
and~(2) at call $t$ give
\begin{align*}
 \Delta_t &= \bigl(c_t(r^{\thr})-f^\ast_{\rho_t}\bigr)-\bigl(c_t(r)-f^\ast_{\rho_t}\bigr)\\
          &\le \bigl(D_{\rho_t}(e_t,\acc_t)-D_{\rho_t}(e_{t+1},\acc_{t+1})\bigr)
              -\bigl(W_{\rho_t}(o_t)-W_{\rho_t}(o_{t+1})\bigr)\\
          &= Z^{\rho_t}_t-Z^{\rho_t}_{t+1}.
\end{align*}
Call $1$ costs the rule nothing, as it has no history and does not reset, so $\Delta_1\le0$.
Summing over the calls and shifting the index of the first sum by one gives
\begin{align*}
 \SchedCost_{r^{\thr}}(S)-\SchedCost_r(S)
   = \sum_{t=1}^{T}\Delta_t
    \le \sum_{t=2}^{T}\bigl(Z^{\rho_t}_t-Z^{\rho_t}_{t+1}\bigr)
   = Z^{\rho_2}_2-Z^{\rho_T}_{T+1}
      +\sum_{t=2}^{T-1}\bigl(Z^{\rho_{t+1}}_{t+1}-Z^{\rho_t}_{t+1}\bigr).
\end{align*}
We bound the three terms separately.

For the first term, before call $2$ we have $e_2=\acc_2=o_2=0$, so
$Z^{\rho_2}_2=D_{\rho_2}(0,0)-W_{\rho_2}(0)=0$ by Lemma~\ref{lem:rate} points~(1) and~(2).

For the second term, $D_\alpha(e,\acc)\ge W_\alpha(e)\ge0$ and $W_\alpha(o)\le K$ by
Lemma~\ref{lem:rate} points~(2) and~(4), so $Z^{\rho_T}_{T+1}\ge-K$.

For the third term, if $a_t=a_{t-1}$ then $\rho_{t+1}=\rho_t$ and the summand is $0$.
Otherwise a boundary lies between calls $t$ and $t+1$, and since $S$ has $P$ runs this
happens for exactly $P-1$ values of $t$. Fix such a $t$, write $\alpha=\rho_t$,
$\beta=\rho_{t+1}$, $(e,\acc)=(e_{t+1},\acc_{t+1})$ and $o=o_{t+1}$.
Since $e\le\acc<K$, Lemma~\ref{lem:rate} point~(4) gives, for every rate $\gamma>0$,
\[
 W_\gamma(e)\le D_\gamma(e,\acc)\le W_\gamma(e)+K.
\]
Together with $W_\gamma(e),W_\gamma(o)\in[0,K]$ and monotonicity in the carry
(Lemma~\ref{lem:rate} points~(2) and~(5)), this yields
\[
 Z^\gamma_{t+1}=D_\gamma(e,\acc)-W_\gamma(o)\in
 \begin{cases}
 [-K,K], & e\le o,\\
 [0,2K], & e>o.
 \end{cases}
\]
The carries $e$ and $o$ are fixed at the boundary, so $Z^\alpha_{t+1}$ and
$Z^\beta_{t+1}$ lie in the same interval of width $2K$.
Hence $Z^\beta_{t+1}-Z^\alpha_{t+1}\le2K$, whatever the direction of the change.

Combining the three bounds gives
\begin{align*}
 \SchedCost_{r^{\thr}}(S)-\SchedCost_r(S)
   \le 0+K+(P-1)\cdot2K = (2P-1)K.
\end{align*}
Since $r$ is arbitrary, this holds in particular for $r=r^\star$, and together with
Proposition~\ref{prop:classical} point~(1) it gives the claim.
\end{proof}

\begin{rem}[Constant growth]\label{rem:base}
With $P=1$, Theorem~\ref{thm:phase} point~(1) says that under constant growth the rule loses
at most $K$ against every schedule. Every rule loses at least $K$. Feed it a constant rate
$\alpha>0$. If it resets, end the session right after its first reset, as the adversary of
ski rental does \cite{karlin1988snoopy}, or after call~$2$ if that reset is before call~$1$,
so that the session has one run. Up to there it has paid the rents of the schedule that
never resets, and at least $K$ more. If it never resets, its cost $g_\alpha(T-1)$ grows
quadratically in $T$, while resetting every $L_\alpha$ calls costs a linear amount, so its
loss exceeds $K$ for $T$ large.
\end{rem}

\begin{rem}[The other discretization]\label{rem:other}
If a reset call pays the price only, the cost of a schedule is
$\SchedCostP_r(S)=\sum_{t=1}^{T}\bigl(r_tK+(1-r_t)h_t\bigr)$. The threshold rule makes the
same decisions under both costs, since its accumulator counts the rent of non-reset calls
only, and the additive bound of Theorem~\ref{thm:phase} holds for $\SchedCostP$ as well:
$\SchedCostP_{r^{\thr}}(S)\le\SchedCostP_r(S)+(2P-1)K$ for every schedule $r$. A reset before call $t$
clears each increment appended since the previous reset exactly once, so
$\SchedCost_r(S)-\SchedCostP_r(S)=\sum_{t\le T}r_th_t$ is $\sum_{t<T}a_t$ minus the carry
$o_{T+1}$ that $r$ leaves after call $T$, and likewise for the rule, whose final carry is
$e_{T+1}$. The proof above bounds $\SchedCost_{r^{\thr}}(S)-\SchedCost_r(S)$ by
$2(P-1)K-Z^{\rho_T}_{T+1}$, so the primed gap is at most
\begin{align*}
    2(P-1)K+\bigl(e_{T+1}-D_{\rho_T}(e_{T+1},\acc_{T+1})\bigr)+\bigl(W_{\rho_T}(o_{T+1})-o_{T+1}\bigr)\le(2P-1)K
\end{align*}
by Lemma~\ref{lem:rate} points~(1) and~(2). The family of Section~\ref{sec:sharp}
ends with rule carry $0$ and a comparator carry that tends to $0$, so the coefficient
$2P-1$ stays sharp. Section~\ref{sec:every} uses $\SchedCost$ only.
\end{rem}

\begin{rem}[Approximate runs]\label{rem:approx}
The bound survives small perturbations of the growth. Let $q$ be a
session with $P$ runs and positive increments, let $\varepsilon\ge0$, and let $S$ have the
same horizon and increments $q_t\le a_t\le(1+\varepsilon)q_t$. Then every schedule $r$ for
$S$ satisfies
\[
 \SchedCost_{r^{\thr}}(S)\le(1+\varepsilon)\bigl(\SchedCost_r(S)+(2P-1)K\bigr)+\varepsilon K.
\]
Write $s$ for the threshold schedule on $S$ and $\theta=\frac{K}{1+\varepsilon}$. For a
fixed schedule, histories and accumulators are linear in the increments with nonnegative
coefficients. So on $q$, the schedule $s$ resets only if its test value is at least
$\theta$, and always if it is at least $K$. Run the proof of point~(1) on $q$, with
$D_\alpha(e,\acc)$ replaced by the largest $V^\alpha_m(e)$ over the indices $m\le k$ with
$\acc+R^\alpha_m(e)\ge\theta$, where $k$ is the crossing index. The proof of
Lemma~\ref{lem:rate} point~(4) uses only $k\le L_\alpha$ and $\acc+R^\alpha_{k-1}(e)<K$, so
it holds for each such $m$, and the boundary and terminal bounds stand. A call without a
reset only removes candidates, so Lemma~\ref{lem:step} point~(1) becomes an inequality in
the right direction. A reset is the index $1$, and the fresh state it leaves is now worth
the largest $V^\alpha_n(0)$ over $n\le L_\alpha$ with $g_\alpha(n)\ge\theta$. This is $0$
for $n=L_\alpha$, and for $n<L_\alpha$ the bound $f^\ast_\alpha\ge\alpha L_\alpha\ge\alpha(n+1)$
gives $V^\alpha_n(0)\le K-g_\alpha(n)\le K-\theta$. Counting the initial state as well,
\begin{align*}
    \SchedCost_s(q)\le\SchedCost_r(q)+(2P-1)K+(K-\theta)(1+N),
\end{align*}
where $N$ is the number of
resets of $s$. Finally, 
\begin{align*}
    \SchedCost_s(S)\le NK+(1+\varepsilon)\bigl(\SchedCost_s(q)-NK\bigr), \quad
    \SchedCost_r(q)\le\SchedCost_r(S) \quad\text{ and }\quad (1+\varepsilon)(K-\theta)=\varepsilon K
\end{align*}
give the claim.
\end{rem}

\section{Sharpness}\label{sec:sharp}

In this section we prove Theorem~\ref{thm:phase} point~(2): for every $\eta>0$ we build a
session with $P$ runs on which the threshold rule loses more than $(2P-1-\eta)K$, so the
charge of $2K$ per change in Section~\ref{sec:phase} is not an artifact of the proof.
At each drop, the threshold rule enters with stale carry while the comparator enters with
none. The slower growth makes the rule pay almost $K$ in repeated rent on that carry before
paying for an extra reset. We end the session at a further threshold reset to retain the
terminal loss of $K$.

\begin{proof}[Proof of Theorem~\ref{thm:phase} point~(2)]
For $P=1$ the argument of Remark~\ref{rem:base}, applied to the threshold rule, gives a loss
of $K$. Let $P\ge2$. We take $K=1$, since multiplying the growth and the price by $K$
multiplies both costs by $K$, and we omit the first call: it has rent $0$, and neither
schedule below resets before it. As in Section~\ref{sec:phase}, call $t\ge2$ has rate
$a_{t-1}$, and run $i$ consists of the calls at rate $\alpha_i$.

We first build the session. Choose an integer $L_1\ge2$ and set
$\alpha_1=\frac{2}{L_1(L_1+1)}$. For $i=2,\dots,P$, define successively
\[
 j_i=\frac{1}{\alpha_{i-1}}-1,\qquad
 L_i=(j_i+1)^2,\qquad
 \alpha_i=\frac{2}{L_i(L_i+1)}.
\]
With incoming carry and accumulator both $\alpha_{i-1}$, $j_i$ calls without new growth
give test value $(j_i+1)\alpha_{i-1}=1$. Choosing $L_i$ much larger than $j_i$ makes
$\alpha_i$ small enough that the
new growth does not cause an earlier reset, as verified below. The choice of $\alpha_i$
also gives $g_{\alpha_i}(L_i)=1$, so $L_i$ is the reset period from an empty state.
These resets happen at exact ties, which the test $\acc_t+h_t\ge K$ counts as resets.
Since $1/\alpha_i=L_i(L_i+1)/2$, all $j_i$ and $L_i$ are integers. Run $i$ has rate
$\alpha_i$ and $L_1+1$ calls if $i=1$, $j_i+L_i+1$ calls if $1<i<P$, and $j_P+L_P$ calls if
$i=P$. As $1>\alpha_1>\alpha_2>\dots>\alpha_P$, the session has exactly $P$ runs.

Next we fix the comparator, the schedule $r$ that resets before the last call of each
nonfinal run and never in run $P$. So it enters every run with carry $0$, and the $n$-th call of run $i$ pays the rent
$n\alpha_i$.

We now follow the threshold rule run by run. In run $1$ the $n$-th call pays $n\alpha_1$ and has test value
$g_{\alpha_1}(n)$, which first reaches $1$ at $n=L_1$: the rule resets there, and the last
call of the run pays $\alpha_1<1$ and does not reset. So the rule leaves run $1$ with carry and
accumulator both equal to $\alpha_1$. Suppose that it enters run $i>1$ with carry and
accumulator $\alpha_{i-1}$. Until its first reset in the run, the $j$-th call pays
$\alpha_{i-1}+j\alpha_i$, so its test value is $(j+1)\alpha_{i-1}+g_{\alpha_i}(j)$. At
$j=j_i$ this is $1+g_{\alpha_i}(j_i)\ge1$, and at $j=j_i-1$ it is
$1-\alpha_{i-1}+g_{\alpha_i}(j_i-1)<1$, because
\begin{align*}
    g_{\alpha_i}(j_i-1)=\frac{(j_i-1)j_i}{(j_i+1)^2((j_i+1)^2+1)}<\frac{1}{j_i+1}=\alpha_{i-1}.
\end{align*}
So the rule first resets before the $j_i$-th call of the run. After that reset the calls pay
$\alpha_i,2\alpha_i,\dots$, so the next reset is $L_{\alpha_i}=L_i$ calls later, before the
$(j_i+L_i)$-th call. In run $P$ this is the last call of the session. In a nonfinal run, one
call is left. It pays $\alpha_i<1$ and does not reset, so the rule leaves run $i$ with carry
and accumulator $\alpha_i$.

Finally, compare the costs on each run. The reset positions and excess costs are summarized
below. Positions are counted within the run, and excess means the rule's cost minus $r$'s.
\begin{center}
\small
\renewcommand{\arraystretch}{1.3}
\begin{tabular}{@{}llll@{}}
\toprule
Run & Rule's resets & $r$'s reset & Excess cost\\
\midrule
First & $L_1$ & last call & $-L_1\alpha_1$\\
Intermediate $i$ & $j_i,\ j_i+L_i$ & last call & $2-3\alpha_{i-1}+\alpha_i$\\
Final $P$ & $j_P,\ j_P+L_P$ & none & $3-\alpha_{P-1}-j_PL_P\alpha_P$\\
\bottomrule
\end{tabular}
\end{center}
For the first row, both schedules reset once and pay the same rents except on the last call,
when $r$ pays $L_1\alpha_1$ more. For an intermediate run, split at the rule's two resets.
The rent differences on the first $j_i$ calls, the next $L_i$ calls, and the last call are
$j_i\alpha_{i-1}$, $-j_iL_i\alpha_i$, and $-(j_i+L_i)\alpha_i$, respectively.
The rule also makes one extra reset. Its excess cost is therefore
\[
 1+j_i\alpha_{i-1}-(j_iL_i+j_i+L_i)\alpha_i
   =2-3\alpha_{i-1}+\alpha_i,
\]
using $j_i\alpha_{i-1}=1-\alpha_{i-1}$ and
$(j_i+1)(L_i+1)\alpha_i=2\alpha_{i-1}$. In the final run the last single call is absent
and the rule makes two extra resets, giving the last row. That row exceeds
$3-3\alpha_{P-1}$, since
$j_PL_P\alpha_P=\frac{2j_P}{(j_P+1)^2+1}<\frac{2}{j_P+1}=2\alpha_{P-1}$. Summing over the runs,
and using $L_1\alpha_1=\frac{2}{L_1+1}$ and $\alpha_i\le\alpha_1=\frac{2}{L_1(L_1+1)}$,
\[
 \SchedCost_{r^{\thr}}(S)-\SchedCost_r(S)\ >\ 2P-1-L_1\alpha_1-3\sum_{i=1}^{P-1}\alpha_i
 \ \ge\ 2P-1-\frac{2}{L_1+1}-\frac{6(P-1)}{L_1(L_1+1)},
\]
which exceeds $2P-1-\eta$ for $L_1$ large. Since $\SchedCost_{r^\star}(S)\le\SchedCost_r(S)$,
this proves part~(2).
\end{proof}

\begin{rem}[Size of the instances]\label{rem:size}
The recursion gives $L_{i+1}=\bigl(\frac{L_i(L_i+1)}{2}\bigr)^2$ and
$\alpha_{i+1}\approx2\alpha_i^4$, so the runs grow doubly exponentially in $P$. This sharpness
construction uses very long sessions. With a bounded horizon or a minimum positive rate, we do not know
the best coefficient.
\end{rem}

\section{The rule or the problem?}\label{sec:every}

Theorem~\ref{thm:phase} is a statement about one rule. Could a different rule, still knowing
neither the runs nor the horizon, pay less for a change of growth? For two runs this section
answers the question completely. Its results fill the second and third rows of
Table~\ref{tab:price}.

\subsection{Decreases: a rule that reacts to them}\label{sec:drop}

As in Section~\ref{sec:sharp}, after a fast run the threshold rule keeps paying the stale
carry of that run during the slow one, and then pays a reset. A rule that sees the decrease
can clear the stale carry at once.

\begin{defn}[Drop-aware rule]\label{def:droprule}
On a session $S$ of horizon $T$, let $h_t$ and $\acc_t$ be as in Definition~\ref{def:thr} for
the schedule being built. Starting from $\acc_1=0$, for $t=1,\dots,T$
\[
(r_t,\ \acc_{t+1})=
\begin{cases}
(1,\ 0) & \text{if } \acc_t+h_t\ge K \text{ or } a_{t-1}<a_{t-2},\\
(0,\ \acc_t+h_t) & \text{otherwise,}
\end{cases}
\]
where $a_{t-1}<a_{t-2}$ is read as false for $t\le2$.
\end{defn}

So the rule is the threshold rule,
except that it also resets at the first call of every run slower than the one before. Since
both tests use only $a_1,\dots,a_{t-1}$, this is a rule, and its schedule on $S$ is the
\textbf{drop-aware schedule} $r^{\mathrm{drop}}$.

\medskip
The analysis of this rule needs one more property of the potentials, which the proof of
Theorem~\ref{thm:phase} did not use: when the growth decreases, the relative value of
every carry can only rise.

\begin{lem}\label{lem:mono}
For every $e\ge0$, the map $\alpha\mapsto W_\alpha(e)$ is nonincreasing on $(0,\infty)$.
\end{lem}

\begin{proof}
Let $0<\alpha\le\beta$. For each $j\ge0$ we show $W_\beta(e)\le V^\alpha_j(e)$, in two cases.

For the first case, $j\le L_\beta$, the minimizing property of $L_\alpha$
(Proposition~\ref{prop:classical} point~(2)) gives
\begin{align*}
    f^\ast_\beta-f^\ast_\alpha
    = f_\beta(L_\beta) - f_\alpha(L_\alpha)
    \ge \frac{g_\beta(L_\beta)-g_\alpha(L_\beta)}{L_\beta}
    = \frac{(\beta-\alpha)(L_\beta+1)}{2}.
\end{align*}
It follows that
\begin{align*}
    V^\beta_j(e)-V^\alpha_j(e) & =  R_j^\beta(e) - R_j^\alpha(e) - j(f_\beta^* - f_\alpha^*) \\
         & = g_\beta(j) - g_\alpha(j) - j(f_\beta^* - f_\alpha^*) \\
         & \leq \frac{(\beta - \alpha) j (j + 1)}{2}- j\frac{(\beta-\alpha)(L_\beta+1)}{2} \\
         &  = \underbrace{(\beta - \alpha)}_{\geq 0}\underbrace{\frac{j (j - L_\beta)}{2}}_{\leq 0} \leq 0.
\end{align*}
So $W_\beta(e)\le V^\beta_j(e) \leq V^\alpha_j(e)$ for every $j \leq L_\beta$.

For $j>L_\beta$, Lemma~\ref{lem:rate} point~(1) gives
\begin{align*}
    W_\beta(e)\le V^\beta_{L_\beta}(e)=L_\beta e\le je\le V^\alpha_j(e) \quad \text{for all } j>L_\beta.
\end{align*}
Hence $W_\beta(e)\le V^\alpha_j(e)$ for every $j\ge0$, and taking the minimum over $j$ gives
$W_\beta(e)\le W_\alpha(e)$.
\end{proof}

\begin{thm}[Drop-aware bound]\label{thm:drop}
Let $K>0$, let $S$ be a session with $P\ge1$ runs, and let $P_-$ and $P_+$ be the numbers of
decreases and of increases between consecutive runs. For every schedule $r$ for $S$, the
drop-aware rule satisfies
\[
 \SchedCost_{r^{\mathrm{drop}}}(S)\;\le\;\SchedCost_r(S)+(1+P_-+2P_+)K.
\]
\end{thm}

\begin{proof}
In this proof, $(e,\acc)$ denotes the drop-aware rule's state.
Telescope as in the proof of Theorem~\ref{thm:phase}. Call a reset made because
$a_{t-1}<a_{t-2}$ a \emph{drop reset}. At a call without a drop reset the rule acts as the
threshold rule, so Lemma~\ref{lem:step} applies, and a boundary at such a call is an
increase, charged at most $2K$ by the jump bound. That bound and the terminal one use
Lemma~\ref{lem:rate} point~(4), which needs $e\le\acc<K$. The drop-aware rule keeps this invariant,
since a drop reset moves it to $(0,0)$. At a decrease from rate $\alpha$ to rate $\beta<\alpha$
the rule makes a drop reset: from state $(e,\acc)$ it pays $e+\beta+K$ and moves to $(0,0)$,
whose potential is $0$ by Lemma~\ref{lem:rate} point~(1). Measure the state before the call with
the old rate and the state after it with the new one. Let $t$ be this call, and let $o$ and
$o'$ be the comparator's carries before and after it. Then $c_t(r)=r_tK+o+\beta$, and the
excess of the call plus the change of $Z$ is
\begin{align*}
 &(e+K-r_tK-o)-W_\beta(o')-\bigl(D_\alpha(e,\acc)-W_\alpha(o)\bigr)\\
 &\qquad=K+\bigl[e-D_\alpha(e,\acc)\bigr]+\bigl[W_\alpha(o)-W_\beta(o)\bigr]
  +\bigl[W_\beta(o)-r_tK-o-W_\beta(o')\bigr].
\end{align*}
The first bracket is at most $0$ by Lemma~\ref{lem:rate} point~(1). So is the second, because $W$ is
nonincreasing in the rate (Lemma~\ref{lem:mono}). The third is at most
$\beta-f^\ast_\beta\le0$ by Lemma~\ref{lem:step} point~(2) at the rate $\beta$, since
$\beta\le\beta L_\beta\le f^\ast_\beta$. So a decrease and the
reset it forces add at most $K$, the boundary included. The terminal term contributes at most
$K$ as before.
\end{proof}

Since $P_-+P_+=P-1$, taking $r=r^\star$ shows that the drop-aware rule loses at most
$(2P-1)K$, as the threshold rule does, and at most $PK$ if every change is a decrease, as on
the family of Section~\ref{sec:sharp}, where the threshold rule loses close to $(2P-1)K$.

\medskip
The drop-aware rule is not a better rule in general. Its additive guarantee is never weaker
than that of Theorem~\ref{thm:phase}, but it pays $K$ at every decrease, whether or not the
decrease leaves stale rent to clear. If the growth alternates between $\alpha$ and
$\alpha-\varepsilon$, with $0<\varepsilon<\alpha<K$, it resets before every other call from
call~$3$ on, at an average cost of $\frac{K+3\alpha-\varepsilon}{2}$ per call, while resetting every
$L_\alpha$ calls costs at most $f^\ast_\alpha<\sqrt{2K\alpha}+2\alpha$ per call. On long
sessions the ratio of its cost to the optimum therefore tends to at least
$\frac{K+3\alpha-\varepsilon}{2\sqrt{2K\alpha}+4\alpha}$, which grows like $\sqrt{\frac{K}{8\alpha}}$ as
$\frac{K}{\alpha}$ grows: unlike the threshold rule, it is not $c$-competitive for any constant $c$.

\subsection{Increases: one reset price}\label{sec:lower}

Increases are different. A rule that loses at most $K$ under constant growth, the least
possible (Remark~\ref{rem:base}), must lose about $2K$ on some session with one increase,
because it cannot tell slow growth that goes on from slow growth that is about to be
followed by a burst.

\begin{thm}[No rule guarantees a loss below $2K$ on at most two runs]\label{thm:lower2}
Let $K>0$, let $\pi$ be a rule and $\eta>0$. There are a session $S$, with one run or with
two runs whose change is an increase, and a schedule $r$ for $S$ such that
\[
 \SchedCost_{r^\pi}(S)\;\ge\;\SchedCost_r(S)+2K-\eta.
\]
\end{thm}
\begin{proof}
Fix an integer $n\ge2$ with $(n-1)\eta\ge6K$, put $\alpha=\frac{3K}{n(n-1)}$ and $\bar a=2K$, and
feed the rule constant rate $\alpha$. If it has not reset within $2n$ calls, end there:
it has paid the rent $g_\alpha(2n-1)$, while resetting once, before call $n+1$, costs
$2g_\alpha(n-1)+n\alpha+K$, which is smaller by $\alpha n(n-1)-K=2K$. Otherwise let $t_1\le2n$ be the first call before which
it resets, and let the increment appended by that call be the burst $\bar a$. The rule, which
has seen only $\alpha$'s, behaves identically on the new session up to that reset. There it pays
$h_{t_1}+K$, with $h_{t_1}\le2n\alpha\le\eta$, and carries the burst into call
$t_1+1$. If it resets again before call $t_1+1$, end the session there: against the
schedule that never resets it has paid two reset prices and saved only the rent $h_{t_1}$.
If it does not, append one more increment $\bar a$ and end after call $t_1+2$: against the
schedule that resets only before call $t_1+1$ it has paid at least as many reset prices,
the burst twice instead of once, and saved only the rent $h_{t_1}$, so it loses at least
$\bar a-\eta$. In both cases the loss is at least $2K-\eta$. In all three cases the increments are
$\alpha$'s followed by none, one or two $\bar a$'s, and $\alpha\le\frac{3K}{2}<\bar a$, so the session
has one run or two runs with an increase.
\end{proof}

So a rule that loses at most $K$ on every session with one run loses at least $2K-\eta$ on a
session with two runs and an increase. In the terms of Table~\ref{tab:price}, this session
forces $b\ge2-\frac{\eta}{K}$ or $b+u\ge2-\frac{\eta}{K}$, so $b+u\ge2$ for every rule. The threshold rule
loses at most $2K$ when the increase is the only change, because the cycle it completes after
the increase costs at most its fair price.

\begin{lem}\label{lem:rise}
Let $0<\alpha\le\beta$ and $m\in\N$ with $g_\alpha(m)<K$, and let $k$ be the crossing index of
$(m\alpha,g_\alpha(m))$ at rate $\beta$. Then
\[
 K+g_\alpha(m)+km\alpha+g_\beta(k)\le mf^\ast_\alpha+kf^\ast_\beta:
\]
the threshold cycle made of $m$ calls at rate $\alpha$ and then $k$ calls at rate $\beta$
costs at most its fair price. Equivalently,
$D_\beta(m\alpha,g_\alpha(m))\le D_\alpha(m\alpha,g_\alpha(m))$.
\end{lem}

\begin{proof}
\noindent\emph{Step 1: locate the crossing indices.}\par\nopagebreak\smallskip

Write $e=m\alpha$, $\acc=g_\alpha(m)$, $L=L_\alpha$, $\Lambda=L_\beta$ and $N=m+k$. Since
$m<L$ and $\acc+R^\alpha_n(e)=g_\alpha(m+n)$, the crossing index at rate $\alpha$ is $L-m$.
Hence
\begin{align*}
 D_\alpha(e,\acc)&=mf^\ast_\alpha-g_\alpha(m),\\
 D_\beta(e,\acc)&=K+g_\beta(k)+km\alpha-kf^\ast_\beta,
\end{align*}
which gives the equivalence.

\smallskip
Since $\acc+R^\beta_\Lambda(e)\ge g_\beta(\Lambda)\ge K$,
we have $k\le\Lambda$. Also $R^\beta_n(e)\ge R^\alpha_n(e)$ gives $k\le L-m$, and
$K\le g_\alpha(m)+km\alpha+g_\beta(k)\le g_\beta(N)$ gives $\Lambda\le N$. Thus
\[
 1\le k\le\Lambda\le N\le L.
\]

\medskip
\noindent\emph{Step 2: expand the potential difference.}\par\nopagebreak\smallskip

Set $\sigma=m\Lambda+kL-L\Lambda=k(N-\Lambda)-(\Lambda-k)(L-N)$.
Substituting $Lf^\ast_\alpha=K+g_\alpha(L)$ and
$\Lambda f^\ast_\beta=K+g_\beta(\Lambda)$ and expanding the quadratic rents gives
\begin{align*}
 L\Lambda\bigl(D_\alpha-D_\beta\bigr)(e,\acc)
 =\Bigl(K-\frac{\alpha L\Lambda}{2}\Bigr)\sigma
    +\frac{\alpha L\Lambda(L-N)(N-\Lambda)}{2}
    +\frac{L\Lambda\,k(\beta-\alpha)(\Lambda-k)}{2}.
\end{align*}
The last two terms are nonnegative. If $\Lambda=L$, then $N=L$ and $\sigma=0$, so the claim
follows. If $\Lambda<L$, then $K>\frac{\alpha L(L-1)}{2}\ge\frac{\alpha L\Lambda}{2}$,
so the claim also follows whenever $\sigma\ge0$.

\medskip
\noindent\emph{Step 3: use the crossing condition when $\sigma<0$.}\par\nopagebreak\smallskip

It remains to treat $\sigma<0$. Its definition forces $(\Lambda-k)(L-N)>0$, hence
$1\le k<\Lambda\le N<L$. Here the crossing condition gives
\[
 \frac{(\beta-\alpha)k(k+1)}{2}\ge K-g_\alpha(N).
\]
Multiply the identity above by $k+1$ and apply this bound to its last term. Since
$L\Lambda(\Lambda-k)>0$, we obtain
\[
 (k+1)L\Lambda\bigl(D_\alpha-D_\beta\bigr)(e,\acc)\ge A(K),
\]
where, for $x\in\mathbb{R}$,
\begin{align*}
 A(x) = (k+1)\Bigl(x-\frac{\alpha L\Lambda}{2}\Bigr)\sigma
    +\frac{(k+1)\alpha L\Lambda(L-N)(N-\Lambda)}{2}
    +L\Lambda(\Lambda-k)\bigl(x-g_\alpha(N)\bigr).
\end{align*}

\smallskip
The function $A$ is affine with nonnegative slope:
\begin{align*}
 (k+1)\sigma+L\Lambda(\Lambda-k)
   =(k+1)k(N-\Lambda)
   +(\Lambda-k)\bigl(L\Lambda-(k+1)(L-N)\bigr)\ge0,
\end{align*}
indeed, $N\ge\Lambda$ and $L\Lambda\ge(k+1)(L-N)$ because $\Lambda\ge k+1$ and $0\le L-N\le L$.

\smallskip
As $K>x_0=\frac{\alpha L(L-1)}{2}$, we get $A(K)\ge A(x_0)$, and a direct computation gives
\begin{multline*}
 A(x_0)=\frac{\alpha L}{2}\Bigl[(k+1)k(N-\Lambda)(2L-1-\Lambda-N)\\
  +(\Lambda-k)(L-N-1)\bigl(\Lambda(L+N)-(k+1)(L-N)\bigr)\Bigr].
\end{multline*}
To check the sign, recall that $1\le k<\Lambda\le N<L$ and that these indices are integers.
Thus $N-\Lambda\ge0$, $\Lambda-k\ge1$, and $L-N-1\ge0$. The two remaining factors satisfy
\begin{align*}
 2L-1-\Lambda-N
   &=(L-N-1)+(L-\Lambda)\ge1,\\
 \Lambda(L+N)-(k+1)(L-N)
   &=(\Lambda-k-1)(L-N)+2\Lambda N\ge0,
\end{align*}
where $\Lambda-k-1\ge0$ follows from $k<\Lambda$. Every factor in the two products is
therefore nonnegative. Since $\alpha L/2>0$, we have $A(x_0)\ge0$, hence $A(K)\ge0$ and
$D_\beta(e,\acc)\le D_\alpha(e,\acc)$.
\end{proof}

\begin{thm}[Two runs]\label{thm:rise}
On a session with at most two runs, the threshold rule loses at most $2K$ if the change is
an increase, and the drop-aware rule loses at most $2K$ in every case. So $2K$ is the smallest
loss a rule can guarantee on such sessions (Theorem~\ref{thm:lower2}). The drop-aware rule
guarantees it, while the threshold rule, which for every $\eta>0$ can lose more than $(3-\eta)K$
(Theorem~\ref{thm:phase} point~(2)), does not.
\end{thm}

\begin{proof}
For an increase from $\alpha$ to $\beta$, the rule starts at $(0,0)$ and each call at rate
$\alpha$ moves $(m\alpha,g_\alpha(m))$ to $((m+1)\alpha,g_\alpha(m+1))$ or, at a reset, to
$(0,0)$. So at the boundary its state is $(m\alpha,g_\alpha(m))$ with $g_\alpha(m)<K$. In the
proof of Theorem~\ref{thm:phase} point~(1) the jump of $Z$ at this boundary is then
$(D_\beta-D_\alpha)(m\alpha,g_\alpha(m))+W_\alpha(o)-W_\beta(o)\le0+K$ by Lemma~\ref{lem:rise} and
Lemma~\ref{lem:rate} point~(2), so the loss is at most $K+K$. On such a session the drop-aware rule
meets no decrease and makes the same decisions. With one run or a decrease,
Theorem~\ref{thm:drop} gives $K$ or $2K$.
\end{proof}

So on two runs the extra loss of the threshold rule comes only from decreases.
Four questions remain open. Does some rule guarantee a loss of at most $PK$ on every
session with $P\ge3$ runs? What is the sharp additive bound for the threshold rule under
repeated increases, where Lemma~\ref{lem:rise} no longer directly applies? Can a rule lose
at most $2K$ on two runs and also be $c$-competitive for some constant $c$?
Is the term $\varepsilon K$ in Remark~\ref{rem:approx} necessary?

\bibliographystyle{acm}
\bibliography{refs}

\end{document}